\documentclass[aps,twocolumn,superscriptaddress]{revtex4-2}
\usepackage{amssymb,amsmath}
\usepackage[colorlinks=true]{hyperref}
\usepackage{graphicx}
\usepackage{color}
\usepackage{tikz}
\usepackage[normalem]{ulem}

\newcommand{\aver}[1]{\left\langle #1 \right\rangle}
\newcommand{\PDF}{{\cal P}}
\newcommand{\rmd}{{\rm d}}
\newcommand{\bs}{\boldsymbol}
\newcommand{\psis}{\psi_{\rm s}}

\begin{document}

\title{Statistics of extreme events in integrable turbulence}

\author{T. Congy}
\email{thibault.congy@northumbria.ac.uk}
\author{G. A. El}
\author{G. Roberti}
\affiliation{Department of Mathematics, Physics and Electrical Engineering, Northumbria University, Newcastle upon Tyne NE1 8ST, United Kingdom}

\author{A. Tovbis}
\affiliation{Department of Mathematics, University of Central Florida, Orlando, Florida 32816, USA}

\author{S. Randoux}
\author{P. Suret}
\affiliation{Univ. Lille, CNRS, UMR 8523 -- PhLAM -- Physique des Lasers Atomes et Mol\'ecules, F-59000 Lille, France}

\date{\today}

\begin{abstract}
  We use  the spectral kinetic theory of soliton gas to  investigate  the likelihood of extreme events in integrable turbulence
  described by the one-dimensional focusing nonlinear Schr\"odinger  equation (fNLSE). This is done by invoking a stochastic  interpretation of the inverse scattering transform for fNLSE and analytically evaluating the kurtosis of the emerging random nonlinear wave field in terms of the spectral density of states of the corresponding  soliton gas. We then apply the general result to two  fundamental scenarios of the generation of integrable turbulence: (i) the asymptotic development of the spontaneous (noise induced) modulational instability of a plane wave, and  (ii) the long-time evolution of strongly nonlinear, partially coherent waves.   In both cases, involving the bound state soliton gas dynamics, the analytically obtained  values of
  the kurtosis are in perfect agreement with those inferred from direct numerical simulations of the the fNLSE, providing the long-awaited theoretical explanation of the respective rogue wave statistics. Additionally,  the evolution of a  particular non-bound state gas is considered providing  important insights related to the validity of the so-called virial theorem.
\end{abstract}

\maketitle

Integrable turbulence (IT) has been introduced by V.~Zakharov~\cite{zakharov_turbulence_2009} as a general theoretical paradigm for the description of random nonlinear  waves in physical systems modeled by integrable equations such as the  Korteweg-de Vries (KdV) or the nonlinear Schrödinger  equation.  Since its inception in 2009, IT has been receiving a growing interest from both theoretical \cite{agafontsev_integrable_2015,agafontsev_integrable_2016,onorato_origin_2016,randoux_nonlinear_2016,roberti_early_2019,agafontsev_extreme_2021}
and experimental \cite{costa_soliton_2014,walczak_optical_2015,suret_single-shot_2016,tikan_single-shot_2018,el_koussaifi_spontaneous_2018,osborne_highly_2019,kraych_statistical_2019,redor_analysis_2020,suret_nonlinear_2020} viewpoints, and has since become a distinct framework to study a large class of complex nonlinear wave phenomena.

Integrable evolution  equations typically arise as leading order approximations of nonlinear dispersive wave systems and often provide a very good description of the core nonlinear dynamics in a variety of physical contexts ranging from water waves to   quantum gases \cite{yang_nonlinear_2010}. The integrable nature of the equations enables analytical solutions  via the inverse scattering transform (IST) method with  both zero and non-zero boundary conditions at infinity \cite{novikov_theory_1984,ablowitz_solitons_1991,yang_nonlinear_2010}. Often seen as a nonlinear generalization of the Fourier method, the IST method consists of three main steps: (i) the direct spectral transform  which  decomposes the scattering  data of the wave field at $t=0$ into the so-called solitonic (related to discrete spectrum) and radiative (related to continuous spectrum) components; (ii) the (simple) time evolution of the scattering data for both components; (iii) the inverse scattering procedure for reconstructing the nonlinear wave-field at $t>0$ from the evolved combined spectral data.

While the above classical, deterministic IST framework has been remarkably successful in many 
problems of nonlinear physics, its ``stochastic'' counterpart addressing the evolution of random initial data in integrable systems (essentially the theory of IT) is still in its infancy, with the majority of physically significant theoretical results being reliant on heavy numerical simulations (see e.g. \cite{agafontsev_integrable_2015,agafontsev_integrable_2016, agafontsev_extreme_2021}) or short-time expansions~\cite{roberti_early_2019}.
The challenge is, given the statistics (the probability density function, the correlations, etc.) of the initial random data for an integrable equation to determine the statistics of the solution at $t>0$.

We propose  a general theoretical approach  to the analysis of the long-time integrable evolution of random wave fields whose typical realizations are dominated by the solitonic spectral component.
We focus on the  statistics of extreme events (a.k.a. rogue waves) in random wave fields developing from certain generic classes of  stochastic initial data for the focusing nonlinear Schr\"odinger equation (fNLSE).
This  fundamental problem of nonlinear  physics has been  the subject of extensive  experimental and numerical investigations for several decades  in various physical contexts \cite{Janssen:03,Onorato:04,  Onorato:05, Nobuhito11, Zhang:14, walczak_optical_2015, Toenger:15,  agafontsev_integrable_2015, randoux_nonlinear_2016, gelash_bound_2019,kraych_statistical_2019, el_koussaifi_spontaneous_2018, suret_single-shot_2016, Narhi:16, tikan_single-shot_2018, dudley2019rogue}.  Recently, statistical estimates for the probability of extreme events have been  derived using large deviations theory \cite{dematteis_rogue_2018,dematteis_experimental_2019}.
However, an exact analytical treatment of the extreme wave statistics remains an open problem.

In this Letter we consider two ubiquitous random waves settings that exhibit extreme events  in the process of the fNLSE evolution: (i) the nonlinear stage of the development of the noise-induced (spontaneous) modulational instability (MI) of a plane wave  \cite{agafontsev_integrable_2015, kraych_statistical_2019}, and (ii) the evolution of the so-called partially coherent waves whose amplitude is given by a slowly varying random function with a given  
statistics \cite{el_koussaifi_spontaneous_2018, suret_single-shot_2016, tikan_single-shot_2018, agafontsev_extreme_2021}.

\begin{figure}[h]
  \centering
  \includegraphics[height=3.9cm]{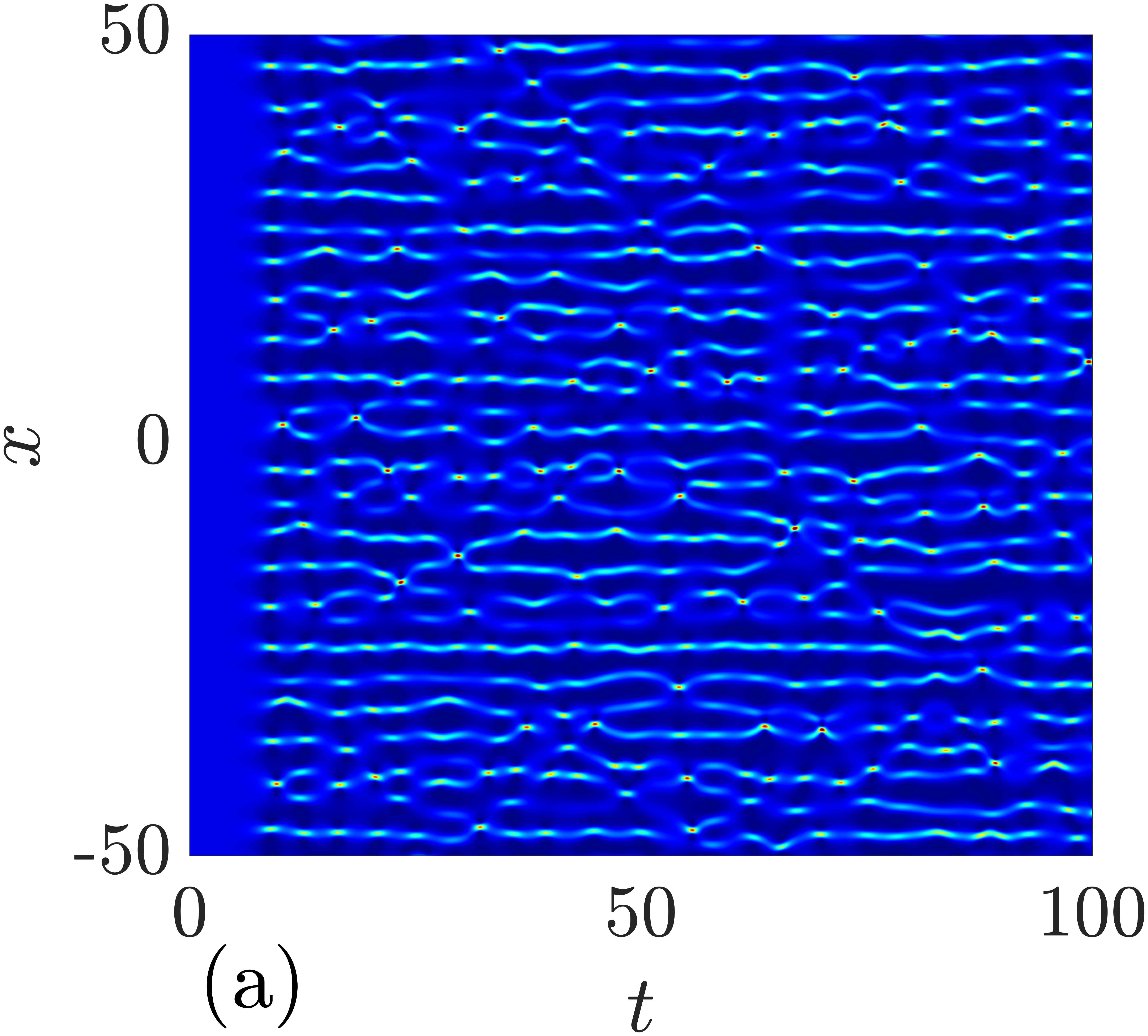}
  \includegraphics[height=3.9cm]{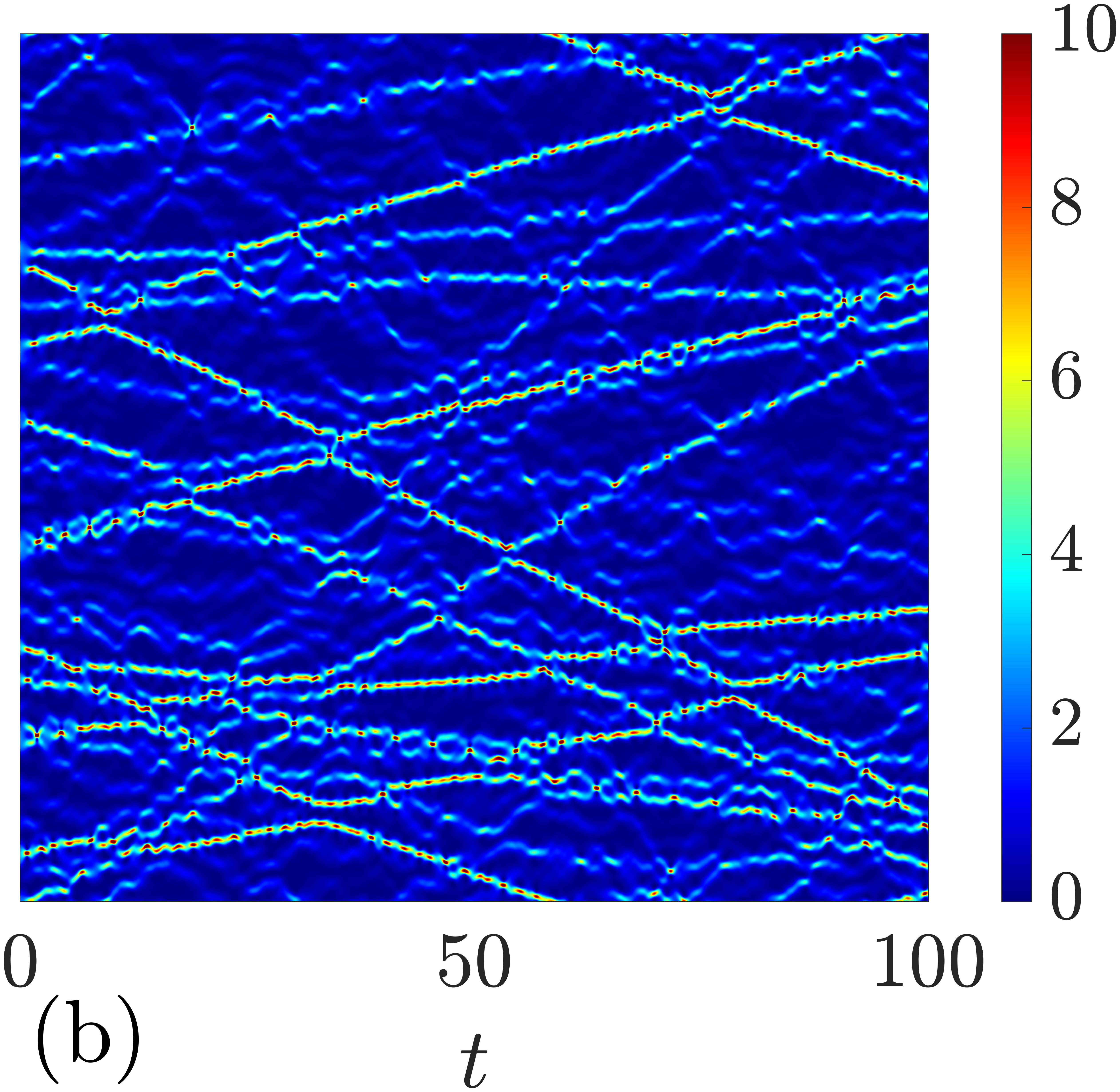}\\[2mm]
  \includegraphics[height=3.6cm]{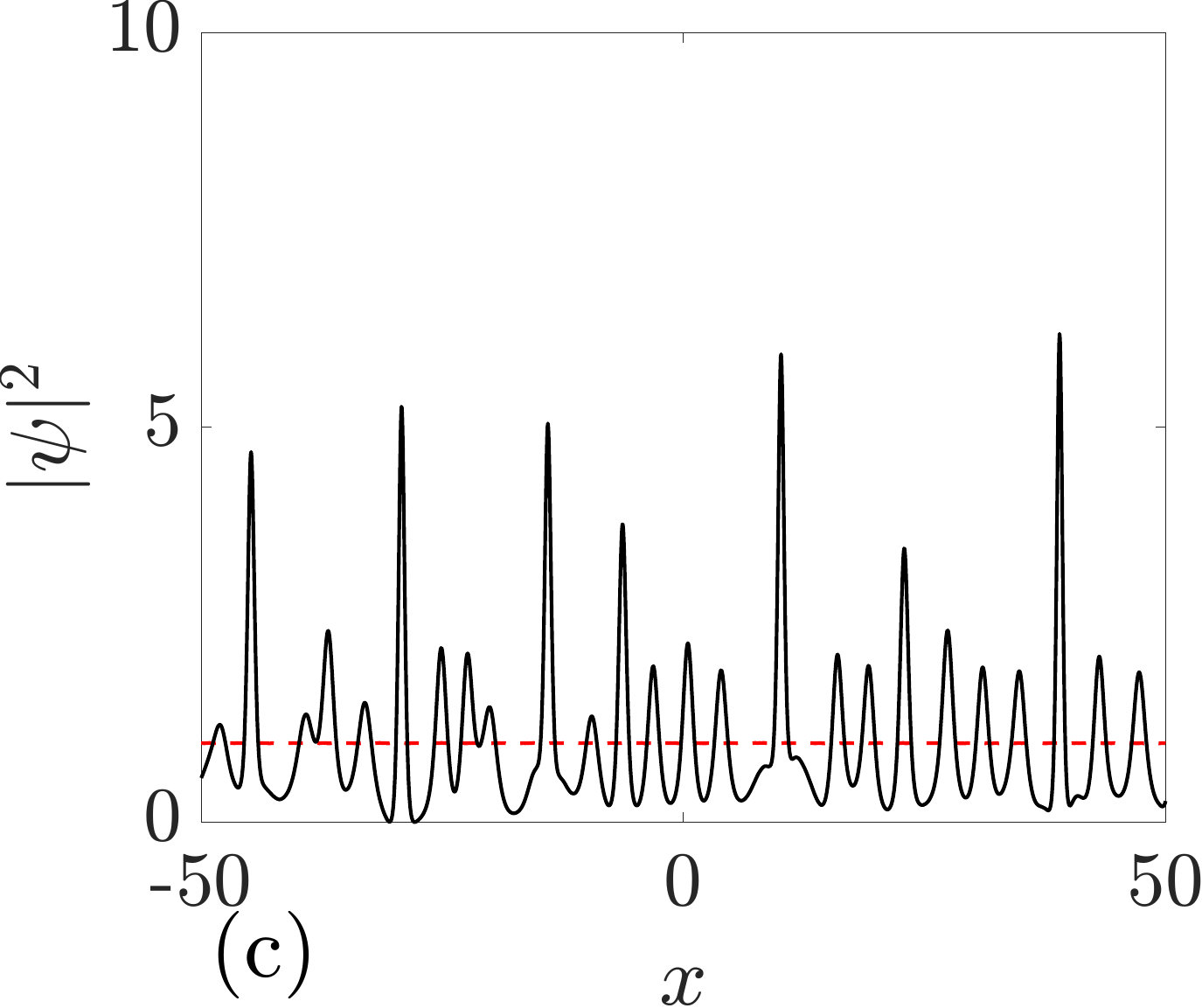}
  \includegraphics[height=3.6cm]{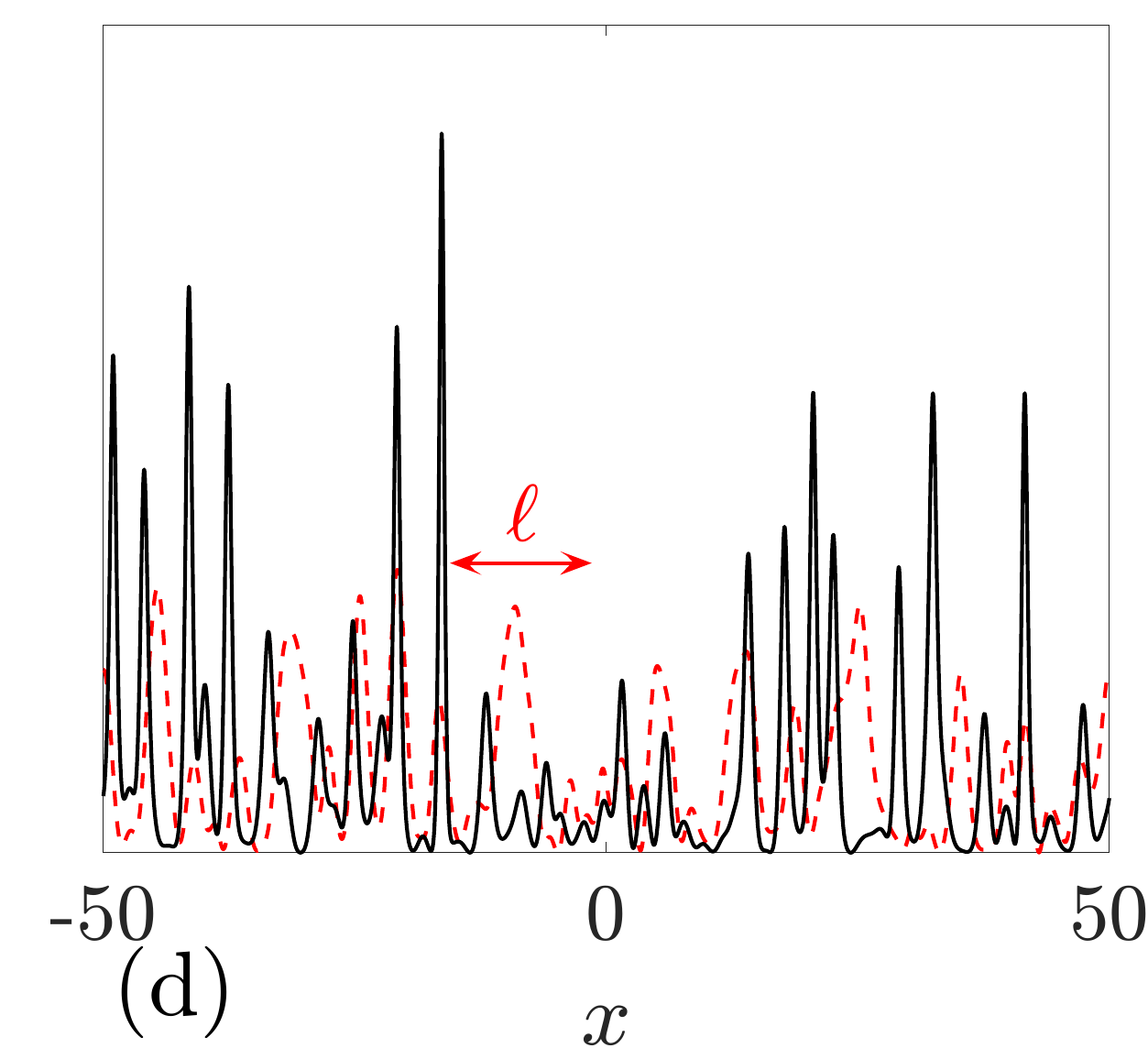}
  \caption{Spatio-temporal plot of $|\psi|^2$ for the asymptotic development of MI (a) and partially coherent waves (b). Snapshots at $t=0$ (dashed red line) and $t=100$ (black solid line) for MI (c) and partially coherent waves (d); $\ell$ represents the typical width of the initial pulses composing the partially coherent wave. The details of the numerical implementation are given in~\cite{gelash_bound_2019,agafontsev_extreme_2021}.}
  \label{fig:example}
\end{figure}

In both settings the amplitude of the initial oscillations, negligible in the MI case and finite in partially coherent waves, dramatically increases with time as depicted in Figure~\ref{fig:example}. The numerical simulations  in \cite{walczak_optical_2015, agafontsev_integrable_2015, agafontsev_extreme_2021} showed that, 
remarkably, the developed IT is characterized by  statistically stationary states  in the long-time regime, but the properties of these states are drastically different for the two types of random input.
This is a clear consequence of  integrability of the wave dynamics exhibiting infinite number of conservation laws,  and in sharp contrast with the properties of classical dissipative hydrodynamic or weak (wave) turbulence  characterized, in the absence of damping or forcing, by  the equipartition of energy and the universal Rayleigh-Jeans Fourier spectra as $t \to \infty$, independently of the initial data \cite{nazarenko_wave_2011}.  In particular, it has been observed that in case (i) the fourth normalized moment $\kappa=\langle |\psi|^4\rangle/|\langle|\psi|^2\rangle^2$ of the probability density function (PDF) of the 
random wave amplitude $\psi$---the kurtosis--- evolves from the initial value of $1$ to $2$ ~\cite{agafontsev_integrable_2015, gelash_bound_2019,kraych_statistical_2019}, while in case (ii) it grows from $2$ to $4$ in the highly nonlinear regime~\cite{agafontsev_extreme_2021}. Although in both cases the kurtosis doubles as a result of the  nonlinear random wave field  evolution, the former case ($\kappa=2$) corresponds to the Gaussian statistics of the asymptotic IT  while the latter ($\kappa=4$) implies  enhanced probability of high amplitude events--- often described as a ``heavy tail'' of the PDF and associated with the rogue wave formation \cite{Janssen:03, Onorato:04, Onorato:05, suret_single-shot_2016, gelash_bound_2019,kraych_statistical_2019, el_koussaifi_spontaneous_2018}. The value of the kurtosis has been derived for partially coherent waves in the {\it weakly nonlinear regime} (i.e. when solitons can be ignored) in the framework of wave turbulence theory ~\cite{Janssen:03}. To the best of our knowledge, there is no theoretical description of the kurtosis doubling in the above scenarios (in the strongly nonlinear regime). Below we present an analytical description of this phenomenon  using recent developments of the spectral theory of {\it soliton gas}.

Soliton gas (SG)
can be seen as an infinite stochastic ensemble of interacting solitons randomly distributed on the whole line \cite{el_soliton_2021}. It
represents  a prominent example of IT that  has been attracting a great deal of interest recently due to the recognition of its ubiquity in various  physical systems~\cite{costa_soliton_2014,Redor:19, Marcucci:19,suret_nonlinear_2020, mossman_dense_2022}.

The theory of SG  was initiated in Zakharov's 1971 paper \cite{zakharov_kinetic_1971} by considering an infinite collection of well-separated (weakly interacting) KdV solitons  randomly
distributed in space  and having some given distribution over the IST spectral parameter $\{ \lambda_k \}$---the discrete spectrum. This  theory  of rarefied SG  has been significantly expanded   by considering dense (strongly interacting) KdV and fNLSE soliton gases within the mathematical framework of the thermodynamic limit for spectral finite-gap solutions and their modulations~\cite{el_thermodynamic_2003, el_kinetic_2005, el_spectral_2020}.
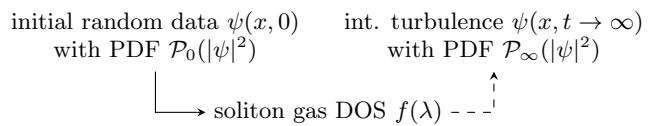
\begin{figure}[h]
  \begin{tikzpicture}

    \node[align=center] at (-4,0)  (1) {initial random data $\psi(x,0)$ \\with PDF $\PDF_0(|\psi|^2)$};
    \node[align=center] at (-1.7,-1)  (2) {soliton gas DOS $f(\lambda)$};
    \node[align=center] at (0.5,0)  (3) {int. turbulence  $\psi(x,t \to \infty)$ \\ with PDF $\PDF_\infty(|\psi|^2)$};

    \draw[-stealth,thin] (1.south) |- (2.west);
    \draw[-stealth,thin,dashed] (2.east) -| (3.south);

  \end{tikzpicture}
  \caption{Stochastic analogue of the nonlinear Fourier (IST) framework for the  fNLSE  with initial data in the form of a random potential dominated by solitonic content}
  \label{fig:IST}
\end{figure}

The key observation is that, in both MI and partially coherent wave settings  
the dynamics are dominated by the solitonic component of the spectrum so that the long-time behavior of the IT can be approximated by appropriate SGs.  In what follows we shall take advantage of the fNLSE  SG theory \cite{el_spectral_2020} to infer important statistical characteristics of the developed, homogeneous IT.  This will be done within the ``stochastic''  version of the IST schematically shown in  Figure~\ref{fig:IST}.  Specifically, one extracts the  spectral statistical distribution---the so-called density of states (DOS)---of the approximating SG from the direct scattering analysis  of random initial data $\psi(x, t= 0)$ (solid line in Figure~\ref{fig:IST}), and then reconstructs the statistics of the long-time asymptotic IT wave field $\psi(x, t \to \infty)$  via the inverse transform of the SG DOS (dashed line) using the invariance in time of the global spectrum statistics for problems with macroscopically homogeneous random initial data.
The time evolution step for scattering data of the traditional IST is replaced by the assumption of an effective stochastization of the soliton phases ensuring spatial uniformity and statistical stationarity of the  SG at $t \to \infty$. Although generally, the direct scattering transform of a random potential represents a complex mathematical problem \cite{pastur_spectra_1992},  we show that the determination of the spectral DOS for the two cases considered here
reduces to evaluating the Abel transform of the PDF of the random initial data.
We then use the relations between the spectral DOS and the fNLSE conserved densities and currents  \cite{tovbis_recent_2022} to determine the kurtosis $\kappa_{\infty}$ of the developed  IT.

\medskip
We consider the fNLSE in the form
\begin{equation}
  \label{eq:nls}
  i \psi_t + \psi_{xx} + 2|\psi|^2
  \psi = 0,\quad \psi(x,t) \in \mathbb{C}.
\end{equation}
The discrete, soliton spectrum $\{\lambda_k\}$ in the linear scattering problem associated with \eqref{eq:nls} lies in complex plane~\cite{zakharov_exact_1972}.
If the
spectrum contains only one
point $\lambda = \xi + i \eta$ in the upper half-plane $\mathbb{C}^+$, the wave field $\psi(x,t)$ is given by the (bright)
soliton solution
\begin{align}
   & \psis(x,t;\lambda) = 2\eta \, {\rm sech} \left[2\eta(x-x_0+4\xi
  t)\right] \nonumber                                                \\
   & \qquad \times \exp \left[-2i \xi (x-x_0) +4i(\eta^2
    - \xi^2) t+ i \sigma_0 \right],
  \label{eq:sol}
\end{align}
where
$x_0,\sigma_0 \in\mathbb{R}$ are constant
parameters describing the soliton spatial position and the phase at $t=0$ respectively;
$2\eta>0$ represents the soliton amplitude, $-4\xi$---the
soliton velocity.  The fNLSE supports $N$-soliton solutions characterized by the spectral set $\{\lambda_k =\xi_k + i \eta_k\}_{k=1}^N$ complemented by the set of $N$ complex norming constants related to the  initial ``positions'' $x_0^k$ and  phases $\sigma_0^k$ of the individual solitons within the $N$-soliton solution \cite{zakharov_exact_1972} (see Supplemental Material \cite{supp} for details). The $N$-soliton solution is called a bound state if all $\xi_k \equiv 0$.

The fNLSE SG can be formally defined via the limit as $N \to \infty$ of $N$-soliton solution with discrete spectrum points $\lambda_k$ distributed with some density  over a domain $\Gamma^+ \subset \mathbb{C}^+$ 
and  appropriately chosen random distributions for the norming constants ensuring certain non-zero spatial density of SG on $\mathbb{R}$.
The key aggregated characteristics of SG  is  the DOS $f(\lambda;x,t)$, defined  as the local density in the spectral phase space $\Gamma^+ \times \mathbb{R}$ so that $f(\lambda;x,t) \rmd \xi \rmd \eta \rmd x$ is the number of soliton states contained in the element $[\xi, \xi+ \rmd\xi]\times [\eta, \eta+ \rmd \eta] \times [x, x+ \rmd x]$ of the phase space at time $t$. For a spatially homogeneous, equilibrium SG $f \equiv f(\lambda)$ globally i.e. $f_t=f_x\equiv 0$.

fNLSE soliton collisions are pairwise, elastic and are accompanied by certain position and phase shifts \cite{zakharov_exact_1972}.  As a result, the effective velocity $s(\lambda)$ of a tracer soliton in a SG is  different from its velocity $-4\xi$ in a ``vacuum'' (free soliton velocity) and is defined by the SG equation of state \cite{el_kinetic_2005, el_spectral_2020}
\begin{equation}
  \label{eq:states}
  s(\lambda) = -4\xi+ \int_{\Gamma^+} \Delta(\lambda,\lambda') f(\lambda') [s(\lambda)-s(\lambda')] |\rmd \lambda'|,
\end{equation}
where the kernel $\Delta(\lambda,\lambda')=\ln | (\lambda^*-\lambda')/(\lambda^*+\lambda')|/{\rm Im}(\lambda)$ describes the asymptotic spatial shift  in a two-soliton collision \cite{zakharov_exact_1972}. 
In this Letter, integrations are written for a 1D  curve $\Gamma^+$. If $\Gamma^+$ is a 2D domain in $\mathbb{C}^+$, the arc integration $\int_{\Gamma^+} \dots |\rmd \lambda|$ should be replaced by $\iint_{\Gamma^+} \dots \rmd \xi \rmd \eta$~\cite{el_spectral_2020}.

The fNLSE has an infinite number of conservation laws
$(p_j[\psi])_t +  (q_j[\psi])_x = 0$ ($j \geq 1$),
see for instance~\cite{yang_nonlinear_2010}. We focus in the
following on
\begin{align}
   & p_1 = |\psi|^2,\quad p_3 = |\psi|^4 -
  |\psi_x|^2,\quad
  q_2 = |\psi|^4 -
  2|\psi_x|^2.
  \label{eq:pq}
\end{align}
$p_3$ is commonly called the energy density, where
$H_{\rm L}=\int |\psi_x|^2 \,\rmd x$ corresponds to the linear kinetic
energy of the physical system, and $H_{\rm NL}=-\int |\psi|^4 \,\rmd x$ to the
nonlinear interaction energy.  It was shown in \cite{tovbis_recent_2022}
that ensemble averages of the densities
$\aver{p_j}$ and currents $\aver{q_j}$ in SGs are given by the moments of
the DOS such that
\begin{align}
   & \aver{p_1} = 4 \,{\rm Im}(\overline{\lambda}),\quad
  \aver{p_3}
  = -\frac{16}{3} \,{\rm Im}(\overline{\lambda^3}),\nonumber \\
   & \aver{q_2} =  4 \, {\rm Im}(\overline{\lambda^2
    s(\lambda)}),
  \label{eq:p1}
\end{align}
where the spectral average  $\overline{h(\lambda)} = \int_{\Gamma^+} h(\lambda) f(\lambda) |\rmd \lambda|$ (see
Supplemental Material \cite{supp} for an alternative derivation of relations \eqref{eq:p1}).
Relations \eqref{eq:p1} are based on ergodicity of a homogeneous SG inherent in the finite-gap construction of ~\cite{el_spectral_2020}, so  $\aver{\dots}$ in \eqref{eq:p1} can be seen as spatial or temporal averages over an infinite period.
A linear combination of these averages yields the value of the kurtosis
for a uniform SG
\begin{equation}
  \label{eq:kurtosis}
  \kappa = \frac{\aver{|\psi|^4}}{\aver{|\psi|^2}^2} =
  -\frac{{\rm Im}(\frac23 \overline{\lambda^3}+ \frac{1}{4}\overline{\lambda^2s(\lambda)})}{{\rm Im}(\overline{\lambda})^2},
\end{equation}
in terms of the spectral DOS.

The special case of the bound state SG is described by the DOS 
$f(\lambda)= \tilde f(\eta) \delta(\xi)$ where $\delta(\xi)$ is the Dirac delta function. Since the corresponding free soliton velocity vanishes, the equation of state \eqref{eq:states} has the solution $s(\lambda)=0$,
and the kurtosis expression~\eqref{eq:kurtosis} simplifies  to
\begin{equation}\label{kappa_BS}
  \kappa = \frac23 \overline{\eta^3} / \overline{\eta}^2,
\end{equation}
where $\overline{\eta^k}$ are moments of the reduced DOS $\tilde f(\eta)$.

\medskip
We now consider the application of the general result~\eqref{eq:kurtosis} to the two fundamental scenarios of the IT development. We first consider the  problem of spontaneous
MI of the plane wave solution $\psi(x,t)=e^{2it}$ of ~\eqref{eq:nls}. At $t=0$ the plane wave
with small random perturbation, a ``noise'' $\phi(x)$:
\begin{equation}
  \label{eq:init_instab}
  \psi(x,0) = 1 + \phi(x),\quad x\in\mathbb{R},
\end{equation}
with $\aver{|\phi|^2} \ll 1$ and
zero average, $\aver{\phi}=0$, where $\aver{\dots}$ stands for the
ensemble average. The plane wave solution  is
unstable with respect to long-wave perturbations which grow
exponentially with time for $t \ll 1$ ~\cite{benjamin_disintegration_1967,zakharov_modulation_2009}. The solution $\psi(x,t)$ develops into an incoherent, strongly oscillating structure, and the wave field statistics becomes stationary in the
long time regime \cite{agafontsev_integrable_2015}.
It was shown in~\cite{gelash_bound_2019}  that the asymptotic dynamics of the spontaneous MI can be accurately modeled by
the uniform, bound state SG with the DOS
\begin{equation}
  \label{eq:condensate}
  \tilde f(\eta) = \frac{\eta}{\pi \sqrt{1-\eta^2}},\quad 0<\eta \leq 1.
\end{equation}

The distribution~\eqref{eq:condensate}, sometimes called the Weyl DOS,
corresponds to the ``soliton condensate''  associated with the
spectral support $ \Gamma^+ =[0, i]$ \cite{el_spectral_2020}.
Substituting the Weyl DOS in~\eqref{kappa_BS}, we obtain
$\kappa \equiv  \kappa_{\infty}=2$ which is twice the initial kurtosis, in perfect agreement with the long-time limit computed numerically  in~\cite{gelash_bound_2019}. This result provides the long-awaited analytical proof of the striking 
statistical property of the spontaneous MI originally discovered in the numerical simulations of \cite{agafontsev_integrable_2015}: the long-time dynamics exhibits large amplitude fluctuations characterized by Gaussian single-point statistics---a counter-intuitive result for strongly nonlinear IT. 

\medskip
We now consider a more general  random initial condition---a ``stochastically modulated'' plane wave---with the following slow variation assumption for the amplitude and phase:
\begin{equation}
  \label{eq:init}
  \frac{\rho_0'(x)}{\rho_0(x)} = { O}(\ell^{-1}),\quad u_0(x) = { O}(\ell^{-1}),\quad \ell \gg 1,
\end{equation}
where   $ \rho_0(x) = |\psi(x,0)|^2$,  $u_0(x)= \partial_x \arg[\psi(x,0)] $,
see e.g.~\cite{roberti_early_2019}.
Additionally, the random process  $\psi(x,0)$  is assumed to be ergodic; an
example of such initial condition is displayed in
Figure~\ref{fig:example}d. The field $\psi(x,0)$ satisfying \eqref{eq:init} is called a  partially coherent wave and can be
regarded as an infinite sequence of broad ``humps'' with random distributions for the
width ${O}(\ell)$ ($\ell \gg 1$), the amplitude ${ O}(1)$, and the position. The randomness of such a wave is realized on the macroscopic scale $L \gg \ell$, while each slowly varying hump corresponds to a coherent
structure (details of the numerical implementation can be found
in~\cite{randoux_optical_2017, roberti_early_2019, Tikan2020Effect}).
At an early evolution time,  each single hump  
exhibits a smooth evolution dominated by nonlinearity \cite{roberti_early_2019} which culminates in the emergence of a gradient catastrophe followed by  the dispersive resolution via an ordered  sequence of coherent structures locally well approximated by the Peregrine breather solutions of fNLSE \cite{Bertola2013Universality, Tikan2017Universality}. Eventually,  at $t \to \infty$, the solution decomposes into a  statistically uniform SG as described below.

In an idealized partially coherent wave with $u_0(x)  \equiv 0$ each hump  can be approximated at leading order by a non-propagating, bound state $N$-soliton solution in the semi-classical limit ($N \to \infty$) \cite{kamvissis_semiclassical_2003,tovbis_semiclassical_2004}.
Within a physically realistic partially coherent wave satisfying~\eqref{eq:init} each soliton  has a small but non-vanishing velocity component $\xi_k \neq 0$ (see \cite{bronski_semiclassical_1996, biondini_spectrum_2023} for precise analytical estimates), as depicted by the trajectories of solitons in Figure~\ref{fig:example}b. However the real part of the soliton spectrum only contributes to a small correction to the
averages~\eqref{eq:p1} and can be neglected in the computation of the wave field statistics.

Since the stochastic process $\psi(x,0)$ is  ergodic, the statistics of the IST spectrum of partially coherent waves  can be
determined from one representative realization of $\psi(x,0)$. In the semi-classical setting $\ell \gg 1$, the spectral distribution of the initial condition for $x \in [0,L]$, is approximated by the Bohr-Sommerfeld density $\varphi_L(\eta) = \int_0^L \frac{\eta}{\pi
\sqrt{\rho_0(x) - \eta^2}} \,\rmd x$ \cite{zakharov_exact_1972}. 
Since the fNLSE evolution is isospectral, the global spectrum statistics is invariant in time and the DOS of the homogeneous SG at $t \to \infty$ is given simply by $\tilde f(\eta) \sim \varphi_L(\eta)/L$ as $L \to \infty$.
Using the change of variable
$x \to \rho=\rho_0(x)$ and the standard geometrical definition of the
PDF 
(see Supplemental Material~\cite{supp} and~\cite{gurbatov_waves_2011}), we obtain that the DOS is given by the Abel transform of the PDF of the field $\rho_0(x)$, denoted~$\PDF_0(\rho)$:
\begin{equation}
  \label{eq:PCW}
  \tilde f(\eta) = \int_{\eta^2}^\infty
  \frac{\eta}{\pi \sqrt{\rho - \eta^2}} \PDF_0(\rho) \rmd\rho,\quad \eta \in  [0,\infty).
\end{equation}
A similar result was derived for the KdV SG in \cite{meiss_drift-wave_1982}. Clearly the  Weyl DOS~\eqref{eq:condensate} for the initial data  in the MI scenario is obtained by taking
$\PDF_0(\rho)=\delta{(\rho - 1)}$, corresponding to the plane wave solution $\psi=1$ at $t=0$.

We can now replace the
average over randomly distributed partially coherent waves by the
average over different realizations of the SG described
by~\eqref{eq:PCW}.  As an illustration, we generate numerically partially coherent waves with Gaussian single-point statistics implying the exponential PDF $\PDF_0(\rho) = \exp(-\rho)$ \cite{nazarenko_wave_2011}. Using formula~\eqref{eq:PCW} we
obtain the Rayleigh distribution $\tilde f(\eta) \sim \eta \exp(-\eta^2) / \sqrt{\pi}$ which yields by \eqref{kappa_BS}  $\kappa_\infty = 4$ in the long-time regime, which is twice the kurtosis at $t=0$. Our theory thus explains the largest value of the asymptotic value  $\kappa_\infty = 4$ observed in the numerical simulations 
performed in the large nonlinearity regime in~\cite{agafontsev_extreme_2021}. Similar to the spontaneous MI scenario, we can infer that the probability of high amplitude waves drastically increases with time for partially coherent waves.

The doubling of the initial kurtosis is 
a general feature
of partially coherent waves in the semi-classical limit, i.e. when the solitons velocity can be neglected to leading order. Indeed~\eqref{eq:PCW} yields a relation between the moments of the SG DOS $\tilde f(\eta)$ and the moments of the initial PDF $\PDF_0(\rho)$,  in particular, $\aver{|\psi(x,0)|^2} = 4 \overline{\eta}$ and $\aver{|\psi(x,0)|^4} = \frac{16}{3} \overline{\eta^3}$ (see Supplementary Material~\cite{supp}).
Thus, the formula \eqref{kappa_BS} derived for the bound state SG  implies that, regardless of the expression for the initial PDF, the kurtosis  of the IT developing as $t \to \infty$  
satisfies
\begin{equation} \label{kappainfty}
  \kappa_{\infty} = 2 \kappa_0,\quad\text{where}\quad \kappa_0 = \frac{\aver{|\psi(x,0)|^4}}{\aver{|\psi(x,0)|^2}^2}.
\end{equation}
An alternative  derivation of \eqref{kappainfty} can be found in the Supplemental Material~\cite{supp}.

\medskip
The general kurtosis formula~\eqref{eq:kurtosis} is also valid for non-bound state SGs ($\xi \neq 0$).  We consider the so-called circular soliton condensate defined in~\cite{el_spectral_2020} by the DOS supported on a semi-circle in the complex plane:
\begin{equation}
  \label{eq:circular}
  f(\lambda) = \frac{{\rm Im}(\lambda)}{\pi} = \frac{\eta}{\pi},\quad \xi^2+\eta^2=1,\quad \eta>0.
\end{equation}
The substitution of the DOS~\eqref{eq:circular} in the equation of state~\eqref{eq:states} yields the effective velocity $s(\lambda) = -8\xi$, which is twice the free soliton velocity, far from the bound state regime.
Now Eqs.~\eqref{eq:kurtosis}, \eqref{eq:circular}
yields $\kappa=2$, similar to the modulational instability induced IT,
which compares very well with the value computed numerically for circular condensates (see Supplemental Material~\cite{supp}).

Although $\kappa=2$ for both the bound state SG generated by MI and the circular soliton condensate, the energy averages $\aver{H_{\rm L}}$ and $\aver{H_{\rm NL}}$ are drastically different for the two SGs.
The average current $\aver{q_2}$ vanishes for bound state SGs (see formula~\eqref{eq:p1}) implying the relation
\begin{equation}
  \label{eq:virial}
  \aver{H_{\rm NL}} = -2\aver{H_{\rm L}},
\end{equation}
i.e. the average interaction energy is twice the average kinetic energy. A dynamical analogue of \eqref{eq:virial} with $\aver{\dots}$ corresponding to a spatial integration and known as the virial theorem, has been previously derived for 2D and 3D fNLSEs using spatial zero boundary conditions \cite{zakharov1986quasi, zakharov1972collapse}. In the bound state SG context, 
one can assume zero boundary conditions for any sufficiently large spatial interval due to the cancellation of the solitons velocity. 
In contrast, we show that $\aver{p_3}=0$  for the circular soliton condensate, yielding the relation $\aver{H_{\rm NL}} = -\aver{H_{\rm L}}$. This does not invalidate the theorem formulated in~\cite{zakharov1986quasi, zakharov1972collapse} since $s(\lambda)\neq 0$ in that case.

\medskip
Summarizing, we have formulated a general theoretical framework for the IST analysis of IT and  have shown  that statistical moments of the long-time development of IT can be effectively computed for certain classes of random initial conditions using the SG approximation. In particular, we have analytically explained the asymptotic doubling of the kurtosis for two ubiquitous nonlinear wave phenomena: the long-time evolution of spontaneous MI~\cite{agafontsev_integrable_2015} and partially coherent waves~\cite{agafontsev_extreme_2021}.  
Concluding, our work paves the way to the determination of the full statistics (i.e. the PDF, the correlations, etc.) in IT and, ultimately, to the realization of the
stochastic IST  schematically shown in  Figure~\ref{fig:IST}.

\begin{acknowledgments}
  This work has been partially supported  by the Agence Nationale de la
  Recherche  through the LABEX CEMPI project (ANR-11-LABX-0007) and the SOGOOD project (ANR-21-CE30-0061), the
  Ministry of Higher Education and Research, Hauts de France council, the
  European Regional Development  Fund (ERDF) through the Nord-Pas de
  Calais Regional Research Council and the European Regional Development
  Fund (ERDF) through the Contrat de Projets Etat-R\'egion (CPER
  Photonics for Society P4S). The authors would like to thank the Isaac Newton Institute (INI) for Mathematical Sciences for support and hospitality during the programme ``Dispersive hydrodynamics: mathematics, simulation and experiments, with applications in nonlinear waves''  when part of the work on this paper was undertaken.  The work of AT, TC and GR was  partially supported by the Simons Fellowships during the INI Programme. GE's and GR's work was also partially supported by EPSRC  Grant Number EP/W032759/1.
  A.T.  was supported in part by the National science Foundation grant DMS-2009647.

  Data Access Statement: this publication is theoretical work that does not require supporting research data.
\end{acknowledgments}

\clearpage
\begin{center}
\textbf{\large Appendix for: Statistics of extreme events in integrable turbulence}
\end{center}
\setcounter{equation}{0}
\setcounter{figure}{0}
\setcounter{table}{0}
\makeatletter
\renewcommand{\theequation}{S\arabic{equation}}
\renewcommand{\thefigure}{S\arabic{figure}}

\section{$N$-soliton solution of fNLSE}
\label{sec:nsoliton}

The numerical modeling of the fNLSE SG is achieved via the construction of $N$-soliton solutions with $N$ large and appropriately chosen  distributions for the discrete spectrum and norming constants.
In our numerics we used  the efficient approach developed in~\cite{gelash_strongly_2018}.

The $N$-soliton solution of the fNLSE
\begin{equation}
  \label{eq:fNLSE}
  i \psi_t+\psi_{xx}+2|\psi|^2\psi =0,
\end{equation}
is given by the ratio of two determinants (see e.g. \cite{novikov_theory_1984, yang_nonlinear_2010})
\begin{align}\label{Nsol}
   & \psi_N(x,t;\boldsymbol{\lambda})= 2i \,\frac{\det \widetilde{M}}{\det M}, \nonumber          \\[2mm]
   & M_{nm} = \frac{e^{\phi_n^*+\phi_m}+e^{-(\phi_n^*+\phi_m)}}{\lambda_n^*-\lambda_m}, \nonumber \\[2mm]
   & \widetilde{M}=\left(\begin{array}{cccc}
                             0             & e^{\phi_1} & \cdots  & e^{\phi_N} \\
                             e^{-\phi_1^*} &            &         &            \\
                             \vdots        &            & M_{n m}              \\
                             e^{-\phi_N^*} &            &
                           \end{array}\right),
\end{align}
where $\phi_k = -i \lambda_k(x-x_0^k)-i \lambda_k^2 t-i \sigma_0^k / 2$.
$\bs \lambda = (\lambda_1,\dots,\lambda_N)$ are the solitons spectral parameters, $x_0^k \in \mathbb{R}$ their initial ``positions'' and $\sigma_0^k \in [0,2\pi)$ their initial phases. Note that $x_0^k$ coincides to the initial position of the soliton with spectral parameter $\lambda_k$ only when the solitons are well separated (e.g. realization of a rarefied SG). It is not the case for realizations of the circular soliton condensate presented thereafter. The norming constants at $t=0$ are
\begin{equation}
  \label{eq:norming}
  C_0^k=\exp(2i \lambda_k x_0^k- i \sigma_0^k).
\end{equation}

If the solution~\eqref{Nsol} is not a bound state
(${\rm Re}(\lambda_j) \neq {\rm Re}(\lambda_k)$ for $j \neq k$), it  asymptotically reduces to a superposition of
well-separated solitons in the limit $|t| \to \infty$~\cite{zakharov_exact_1972}
\begin{align}
  \label{eq:sum}
  \psi_N(x,t;\boldsymbol{\lambda}) \sim \sum_{k=1}^N \psi_{\rm s}(x,t;\lambda_k,x_\pm^k,\sigma_\pm^k),
\end{align}
where $\psi_{\rm s}$ is the one-soliton solution:
\begin{align}
   & \psis(x,t;\lambda,x_0,\sigma_0) = 2\eta \, {\rm sech} \left[2\eta(x-x_0+4\xi
  t)\right] \nonumber                                                             \\
   & \qquad \times \exp \left[-2i \xi (x-x_0) +4i(\eta^2
    - \xi^2) t+ i \sigma_0 \right],
  \label{eq:sol}
\end{align}
with $\lambda=\xi+i\eta$, and where $x_\pm^k,\sigma_\pm^k$ are the positions and phases of the $k$-th soliton at $t \to \pm \infty$. Because of the interaction between the $N$ solitons, occurring at finite time $t$, the position $x_-^k$ and phase $\sigma_-^k$ are different from the position $x_+^k$ and phase $\sigma_+^k$.

In numerical applications (e.g. realizations of the circular condensate), $\psi_N(x,t;\bs \lambda)$ can be evaluated efficiently using the dressing method presented in~\cite{gelash_strongly_2018}. The numerical scheme is subject to roundoff errors during summation of exponentially small and large values for a large number of solitons $N$, and the implementation of high precision arithmetic routine is necessary to generate solutions with the number of solitons $N>10$. We note that the fNLSE considered in \cite{gelash_strongly_2018} has a different normalization than \eqref{eq:fNLSE}, and one should substitute $t$ by $2t$ in the dressing method to obtain the $N$-soliton solution of \eqref{eq:fNLSE}.

\section{Numerical implementation of random initial conditions}

The evolution of random initial conditions, noise (MI) and partially coherent wave, displayed in Figure~1 in the main text is obtained via direct numerical resolution of~\eqref{eq:fNLSE} with periodic boundary conditions: $\psi(-L/2,t) = \psi(L/2,t)$.
The corresponding initial value problem is solved via a pseudo-spectral, adaptive fourth order Runge-Kutta method with the spatial discretization $\Delta x = L/N_x$, $N_x=2^{12}$.

Both the noise and the partially coherent wave is implemented with a sum of incoherent, discrete Fourier components:
\begin{equation}
  \Psi_{\rm RP}(x) = \Psi_0 \sum_{j=1}^{N_{\rm modes}} \exp\left[-\frac{k_j^2}{2\Delta k^2}+i (k_jx+\sigma_0^j)\right],
\end{equation}
with $k_j = \frac{2\pi}{L} j$ and $\sigma_0^j$ uniformly distributed between $0$ and $2\pi$. The initial conditions implemented to generate Figure~1 in the main text are:
\begin{itemize}
  \item Noise: $\psi(x,0) = 1+ \Psi_{\rm RP}(x)$ with $L=200$, $N_{\rm modes}=4096$, $\Delta k=11.3$, and $\Psi_0=2.4\times 10^{-4}$.
  \item Partially coherent wave: $\psi(x,0) = \Psi_{\rm RP}(x)$ with $L=100$, $N_{\rm modes}=4096$ and $\Delta k=11.3$. We choose $\Psi_0$ to obtain $\frac1 L \int_{-L/2}^{L/2} |\psi(x,0)|^2 \rmd x=1$; in particular we have the initial PDF ${\cal P}_0(\rho=|\psi|^2) = \exp(-\rho)$, see for instance~\cite{nazarenko_wave_2011}.
\end{itemize}
Both initial conditions develop into realizations of spatially uniform SGs in the long time regime.
The numerical value of the kurtosis is obtained by averaging solutions of the initial value problem (with different randomly distributed initial conditions) in the long time regime. The averaging procedure corresponds to an ensemble average, over the different realizations, as well as a spatial average between $x=-L/2$ and $x=L/2$. 

The evolution of the kurtosis in time has been precisely determined numerically in~\cite{agafontsev_integrable_2015} for MI and in~\cite{agafontsev_extreme_2021} for partially coherent waves. 
In the first case, after an initial exponential growth of the noise due to MI (similar to evolution depicted in Figure~1a in the main text), the value of the kurtosis oscillates around the expected value $\kappa_\infty=2$.  Authors in~\cite{agafontsev_integrable_2015} have observed in numerical simulations that the amplitude of the oscillations decays with time as $t^{-3/2}$ and a stationary regime establishes in the long time regime.
In the second case, the stationary regime established after a finite time $t=O(\varepsilon^{-1})$ where $\varepsilon$ is the dimensionless, small parameter characterizing the semi-classical limit. Since the typical amplitude of the partially coherent waves is $O(1)$ in this work, the semi-classical parameter is simply given by $\varepsilon \sim \ell^{-1}$ where $\ell$ is the typical width of the humps.

Since the problem implemented numerically has periodic boundary conditions, one could expect Fermi-Pasta-Ulam-Tsingou (FPUT) recurrence phenomenon
for which one realization $\psi(x,t)$ evolves back to the initial condition $\psi(x,0)$. In that sense, the initial value problem solved numerically does not reach an asymptotic state described by stationary statistics. The stationarity is achieved for $L \to \infty$ when the recurrence time also goes to infinity. 
In practice, the numerical simulations are performed with $L=O(10^2)$ such that a ``quasi-stationary'' regime establishes well before the FPUT recurrence; the procedure is detailed in the reference~\cite{agafontsev_integrable_2015}. Note that we do not observe FPUT recurrence numerically in the range of parameters considered in this work.

\section{Circular soliton condensate}

The DOS of the circular soliton condensate is given by formula (13) in the main text \cite{el_spectral_2020}:
\begin{equation}
  \label{eq:support}
  f(\lambda) = \frac{{\rm Im}(\lambda)}{\pi},\; \Gamma^+ = \{\lambda \in \mathbb{C} \;|\; |\lambda|=1,\; {\rm Im}(\lambda)>0 \},
\end{equation}
The effective velocity is $s(\lambda)=-8{\rm Re}(\lambda)$.
The spectral parameter $\lambda \in \Gamma^+$ can be parametrized by an angle: $\lambda = \xi + i \eta$ with $\xi = \cos\theta$ and $\eta = \sin\theta$ such that the arc integration $\int_{\Gamma^+} \dots f(\lambda)|\rmd \lambda|$ becomes $\int_0^\pi \dots \sin(\theta)\rmd \theta/\pi$. Equations~(4) and (5) from the main text yield:
\begin{align}
   & \aver{|\psi|^2}= \frac{4}{\pi} \int_0^\pi \sin^2\theta \,\rmd \theta = 2,                                      \\
   & \aver{|\psi|^4 - |\psi_x|^2} \nonumber                                                                         \\
   & \hspace{1cm}= \frac{16}{3\pi} \int_0^\pi \left( \sin^4\theta- 3\cos^2\theta\sin^2\theta\right) \rmd \theta= 0, \\
   & \aver{|\psi|^4 -
    2|\psi_x|^2} = -\frac{64}{\pi} \int_0^\pi \cos^2\theta\sin^2\theta \,\rmd \theta =-8,
\end{align}
yielding the value of the kurtosis $\kappa =2$.

The special SG termed  in \cite{el_spectral_2020}  the ``circular soliton condensate'' is implemented numerically using the $N$-soliton solution with $N\gg 1$ and the spectral parameters $\lambda_k$ distributed in a certain way along a circle in the complex plane. Since the gas of interest has stationary statistical properties, we distribute the spectral parameters and norming constants
such that $\psi_N(x,t;\bs \lambda)$ approximates a typical realization of the SG in a finite region of space $x \in [-L,L]$; we set here $t=0$, such that the norming constants are given by~\eqref{eq:norming}. Note that the homogeneous SG is considered here as  established at the outset, and does not result from the long time evolution of random initial conditions as depicted in Section II.

\begin{figure}
  \includegraphics{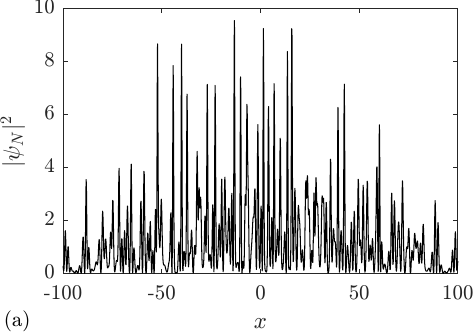}\\[5mm]
  \includegraphics[scale=0.333]{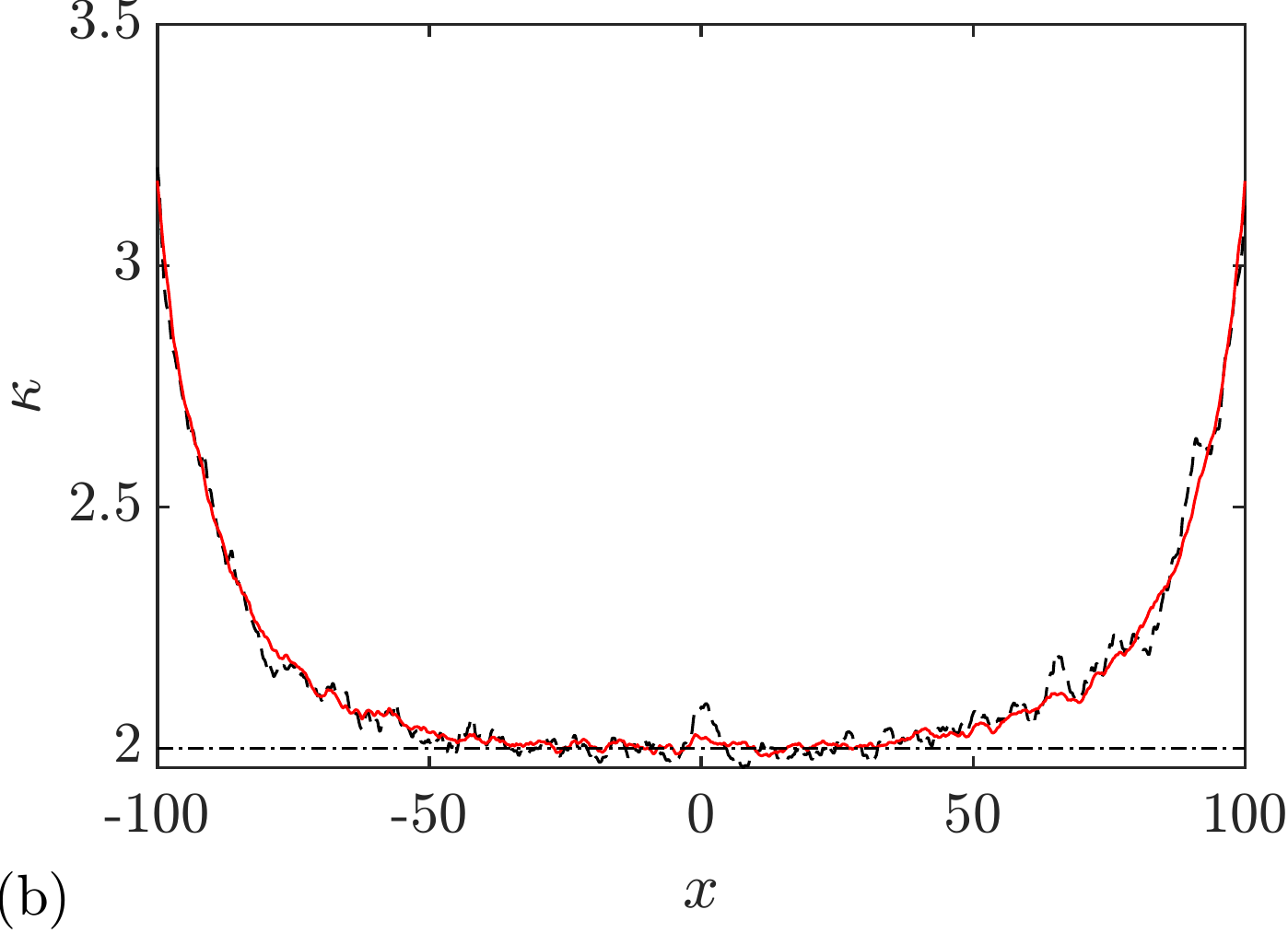}
  \caption{(a) Numerical realization $\psi_N(x,0;\boldsymbol \lambda)$ of the  fNLSE circular soliton condensate with $N=100$. (b) Kurtosis obtained by averaging $9000$ $N$-soliton solutions; the black dashed line (red solid line) corresponds to the choice $\alpha=0.07$ ($\alpha=0.15$).}
  \label{fig:circular}
\end{figure}

The discrete spectral parameters $\lambda_k$ of the approximating $N$-soliton solution \eqref{Nsol} are sampled from the continuous distribution on $\Gamma^+$ with the normalized density $f(\lambda)/\kappa$ where
\begin{equation}
  \label{eq:rho}
  \kappa = \int_{\Gamma^+} f(\lambda) |\rmd \lambda| = \int_0^\pi \frac{\sin \theta}{\pi} \rmd \theta= \frac{2}{\pi},
\end{equation}
corresponds to the spatial density of solitons, i.e. the number of solitons per unit interval of  $x \in \mathbb{R}$.
Norming constants' norm $|C_0^k|$ and argument $-\sigma_0^k$ are uniformly distributed in the intervals $[1-\alpha,1+\alpha]$ and $[0,2\pi)$ respectively, with $\alpha < 1$. A typical $N$-soliton solution with $N=100$ and $\alpha=0.15$ is shown in Figure~\ref{fig:circular}a. If $\alpha$ is too small, solitons accumulate at the position $x=0$ and the SG statistics is no longer uniform; in particular the $N$-soliton solution is symmetric if $\alpha=0$~\cite{gelash_bound_2019}. In practice we choose a value of $\alpha$ such that the SG statistics is uniform in the neighborhood of $x=0$.

Moments of the wave field  $\aver{h[\psi]}$ are obtained numerically with a realization averaging and a local, spatial averaging of the $N$-soliton solutions
\begin{equation}
  \aver{h[\psi(x,t)]} = \frac{1}{M} \sum_{j=1}^M \frac{1}{\ell} \int_{x-\ell/2}^{x+\ell/2} h[\psi_N(y,t;{\bs \lambda}_j)] \rmd y,
\end{equation}
where $M$ is the number of ``realizations''. Figure~\ref{fig:circular}b displays the variation of the kurtosis computed over $M=90000$ $N$-soliton solutions ($N=100$), and a local spatial average with $\ell = 30$ for $\alpha \in \{0.07,0.15\}$. If $\alpha=0.15$, the kurtosis is approximately uniform in the region $x\in[-50,50]$ with $L=50$, and we can assume that $N$-soliton solutions describe realizations of the circular SG in the region $[-L,L]$. This assumption is confirmed by the agreement between the kurtosis determined analytically $\kappa=2$, and the kurtosis evaluated numerically in the region $[-L,L]$ which is approximately equal to $2$.

The increase of the kurtosis above $2$ in the neighborhood of $x=0$ for a smaller value of $\alpha$ is a numerical artifact of the implementation as explained above, the kurtosis decreasing then to theoretical value $2$ for $|x|> 5$. The kurtosis also increases outside the region $[-L,L]$ as the soliton density decreases, and the $N$-soliton solutions no longer represent realizations of the dense, circular soliton condensate.

\section{Averages of densities and fluxes}

In this section we compute  ensemble averages of the densities $p_j[\psi]$ and currents $q_j[\psi]$
for a homogeneous SG with DOS $f(\lambda)$. A similar physical derivation for defocusing NLSE and Kaup-Boussinesq SGs is presented in~\cite{congy_soliton_2021}. The nonlinear wave field in a homogeneous SG
represents an ergodic random process, both in $x$ and $t$, and the
ergodicity property implies that ensemble averages in the SG
can be replaced by the corresponding spatial or temporal averages over
one realization $\psi(x,t)$. In particular, we have
\begin{align}
   & \aver{p_j[\psi]} = \lim_{L \to \infty} \frac 1 L
  \int_{0}^{L} p_j[\psi(x,t)] \,\rmd x,               \\
   & \aver{q_j[\psi]}= \lim_{T \to \infty}\frac 1 T
  \int_{0}^{T}  q_j[\psi(x,t)] \,\rmd t,
\end{align}
where $\psi(x,t)$ is a single representative realization of the
SG.

\begin{figure*}
  \includegraphics{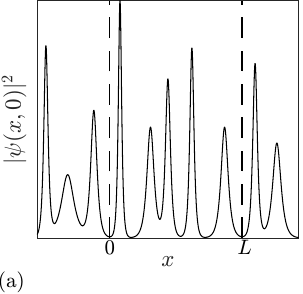}\hspace*{5mm}
  \includegraphics{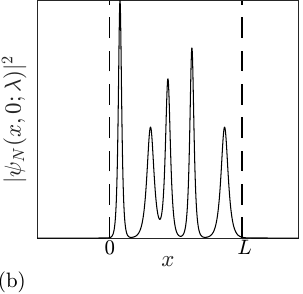}\hspace*{5mm}
  \includegraphics{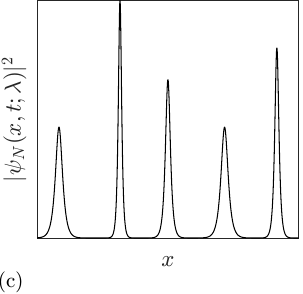}
  \caption{(a) Typical realization $\psi(x,0)$ of a fNLSE SG. (b) Approximation of $\psi(x,0)$ for $x\in [0,L]$ by a (non-bound state)$N$-soliton solution $\psi_N(x,0;\boldsymbol \lambda)$. (c) $N$-soliton solution at a time $t \gg 1$.}
  \label{fig:explosion}
\end{figure*}

We define
\begin{equation}
  A_j = \int_0^L p_j[\psi(x,0)] \rmd x,
\end{equation}
where $L \gg 1$, such that
\begin{equation}
  \label{eq:aver1}
  \aver{p_j[\psi]} = \frac{A_j}{L} + O(L^{-1}).
\end{equation}
Note that the average is evaluated at $t=0$ for computational
convenience only, and the ergodicity property of the gas guarantees that
this average is constant in time. We define the $N$-soliton solution $\psi_N(x,t;\bs \lambda)$ (cf. Section~\ref{sec:nsoliton}), which approximates $\psi(x,t)$ in the spatial window $[0,L]$ at time $t=0$ and vanishes  outside the window:
\begin{equation}
  \label{eq:window0}
  \psi_N(x,0;\bs \lambda) \sim
  \begin{cases}
    \psi(x,0), & \text{if } 0  \le x \le L, \\
    0,         & \text{otherwise.}
  \end{cases}
\end{equation}
Comparison between $\psi(x,0)$ and $\psi_N(x,0;\bs \lambda)$ is displayed in Figure~\ref{fig:explosion}.
By construction, the spectral content of the realization $\psi(x,t)$ is purely solitonic, which guarantees that the truncated portion of SG~\eqref{eq:window0} can be approximated by  $N$-soliton solution for $L$ sufficiently large.
Spectral parameters $\lambda_k$ of the $N$-soliton solution  are distributed by $L f(\lambda)$. Since $L \gg 1$ and $f(\lambda)=\mathcal{O}(1)$, the total number of solitons $N$ is large.
Since $\psi_N(x,0;\bs \lambda)$ exponentially decays in the regions $x<0$ and $x>L$, we have
\begin{equation}
  \label{eq:In}
  A_j
  \sim \int_{-\infty}^{+\infty} p_j[\psi_N(x,0;\bs \lambda)] \rmd x.
\end{equation}
$\psi_N$ is only an approximation of the realization $\psi$ in $[0,L]$, especially at the boundaries $x=0$ and $x=L$ where $\psi_N(x,0;\bs \lambda)$ decays to $0$, see Figure~\ref{fig:explosion}b, and the correction to the integral~\eqref{eq:In} is $O(1)$.
The Kruskal integrals
$\int_\mathbb{R} p_j[\psi_N(x,t;\bs \lambda)] \rmd x$
are time-conserved quantities and
\begin{equation}
  A_j \sim \int_{-\infty}^{+\infty} p_j[\psi_N(x,t;\bs \lambda)] \rmd x,
\end{equation}
for any time $t$.

We suppose first that the $N$-soliton solution is not a bound state. In the limit $t \gg 1$, the
$N$-soliton solution can be approximated by a linear superposition of
distinct solitons~\eqref{eq:sum} (cf. Figure~\ref{fig:explosion}c), and $A_j$ is given
by
\begin{align}
  \label{eq:sumI}
  A_j \sim \sum_{k=1}^N P_j(\lambda_k),\quad
  P_j(\lambda_k) =  \int
  p_j[\psis(x,t;\lambda_k)] \rmd x,
\end{align}
where $\lambda_k$ are distributed by the spectral density $L f(\lambda)$, and $\psis$ is the one-soliton solution~\eqref{eq:sol}.
We have in particular:
\begin{equation}
  P_1(\lambda_k) = 4{\rm Im}(\lambda_k),\quad
  P_3(\lambda_k) = -\frac{16}{3} {\rm Im}(\lambda_k^3).
\end{equation}
The continuous limit of~\eqref{eq:sumI} for a sufficiently large number
of solitons $N$ reads
\begin{equation}
  \label{eq:continuousI}
  A_j \sim \int_{\Gamma^+} P_j(\lambda) \, L f(\lambda) \,
  |\rmd \lambda|,
\end{equation}
where integration occurs over the spectral support of $f(\lambda)$.
Substituting the integral~\eqref{eq:continuousI} in the limit~\eqref{eq:aver1}, we obtain
\begin{align}
  \label{eq:p1}
   & \aver{p_1[\psi]} =
  4 \, {\rm Im}\left(\int_{\Gamma^+} \lambda f(\lambda) |\rmd \lambda| \right), \\
  \label{eq:p3}
   & \aver{p_3[\psi]}
  = -\frac{16}{3} {\rm Im}\left(\int_{\Gamma^+} \lambda^3 f(\lambda) |\rmd \lambda|\right).
\end{align}

\begin{figure*}
  \includegraphics[scale=0.333]{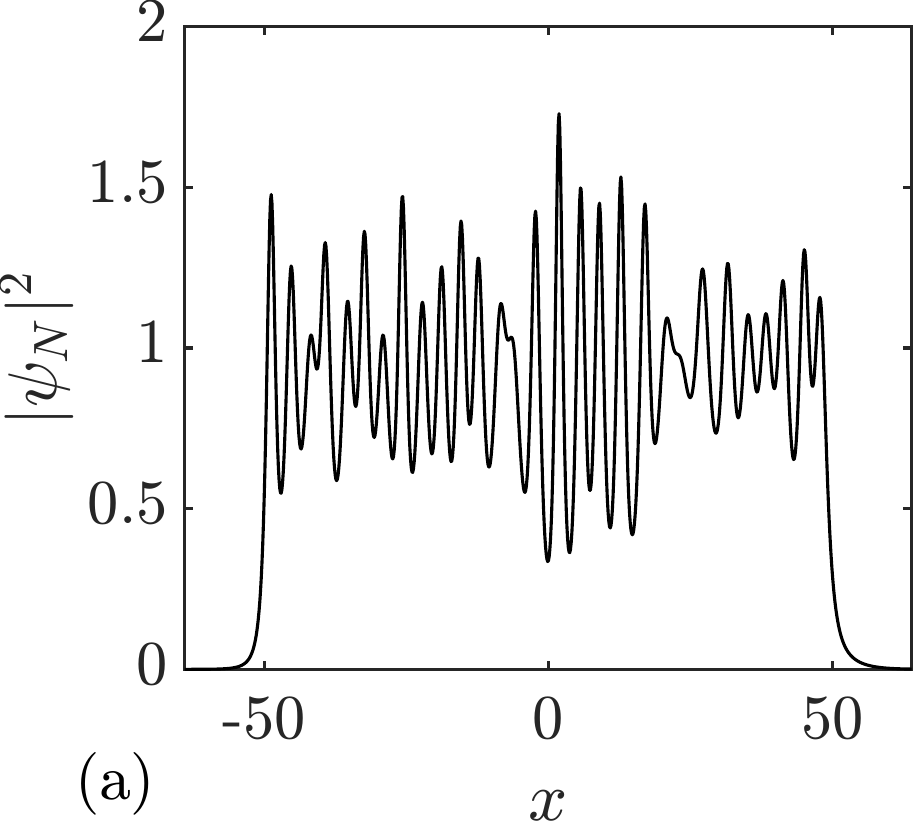}\hspace*{5mm}
  \includegraphics[scale=0.333]{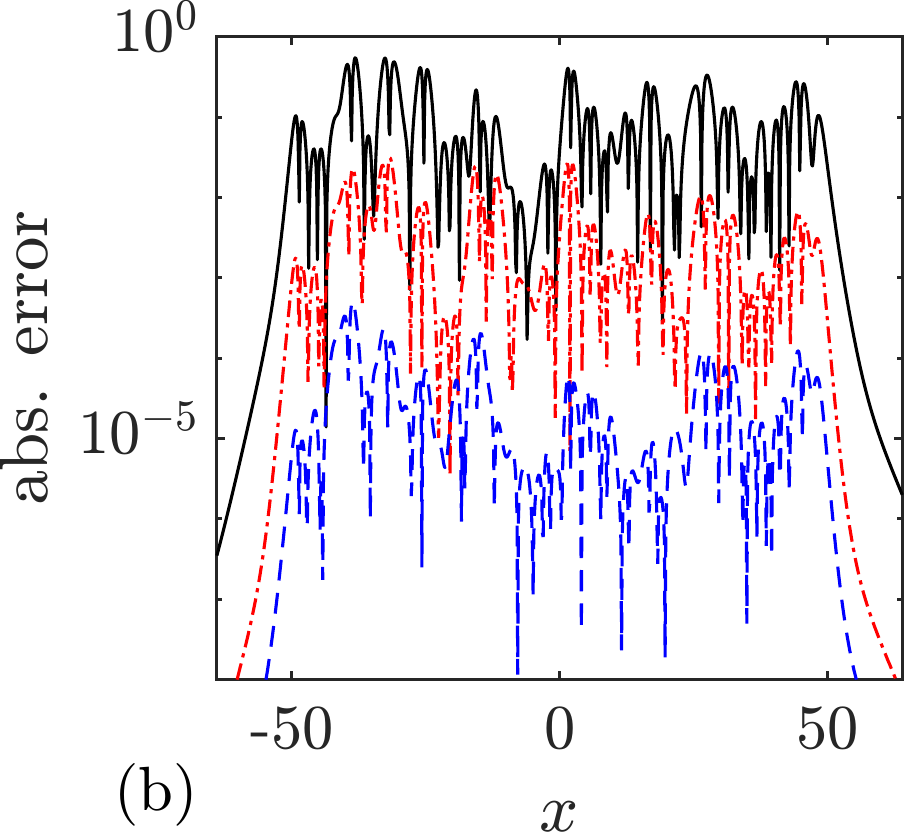}\hspace*{5mm}
  \includegraphics[scale=0.333]{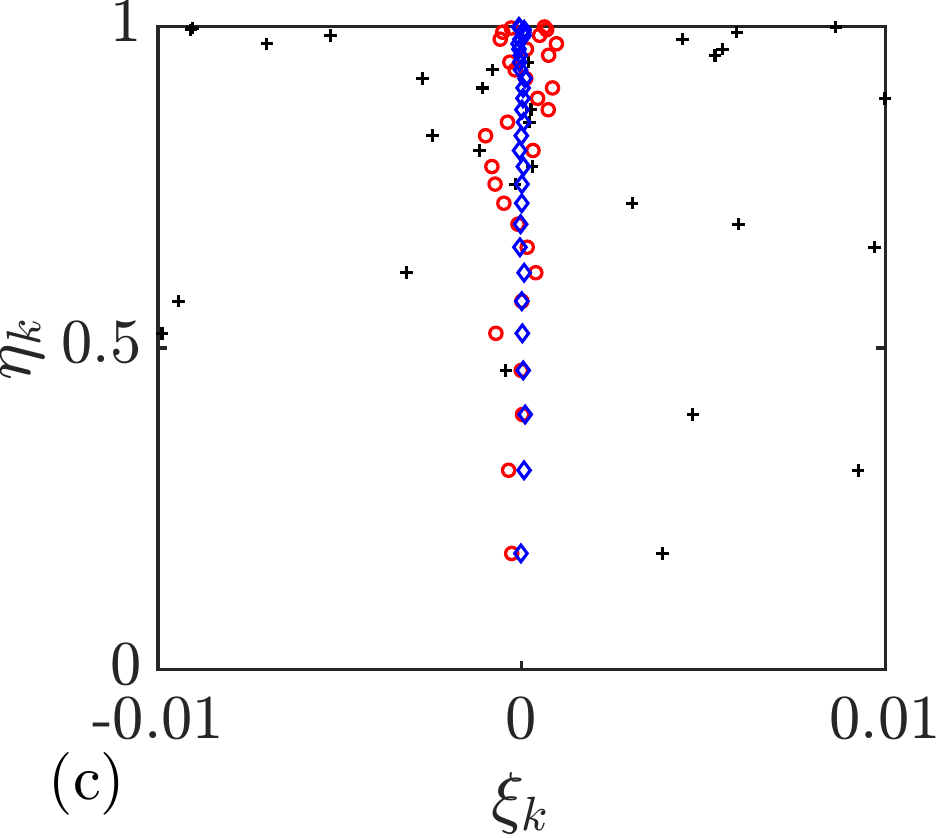}
  \caption{$N$-soliton solution with $N=30$. $\eta_k={\rm Im}(\lambda_k)$ are distributed with the Weyl distribution (Equation (9) in the main text), and $\xi_k={\rm Re}(\lambda_k)$ are uniformly distributed in $[-\varepsilon,\varepsilon]$; $x_0^k=0$ and $\sigma_0^k$ are uniformly distributed in $[0,2\pi)$ (a) $N$-soliton solution for $\varepsilon=0$. (b) Absolute error $|\psi_N(x,0;\boldsymbol{\lambda}^\varepsilon)|^2-|\psi_N(x,0;\boldsymbol{\lambda})|^2$ with $\varepsilon=10^{-2}$ (black solid line), $\varepsilon=10^{-3}$ (red dash-dotted line) and $\varepsilon=10^{-4}$ (blue dashed line). (c) Spectrum $\lambda_k^\varepsilon=\xi_k^\varepsilon+\eta_k$ of the $N$-soliton solution with $\varepsilon=10^{-2}$ (black markers), $\varepsilon=10^{-3}$ (red circles) and $\varepsilon=10^{-4}$ (blue diamonds).}
  \label{fig:vto0}
\end{figure*}

Suppose now that the $N$-soliton solution is a bound state with ${\rm Re}(\lambda_k)=0$ for all $k$ without loss of generality.
Define $\psi_N(x,t;\bs \lambda^\varepsilon)$ a ``copy'' of the $N$-soliton
solution $\psi_N(x,t;\bs \lambda)$ having the same distribution of ${\rm Im} (\lambda_k)$ but with ${\rm Re}(\lambda_k) \equiv \, {\rm Re}(\lambda^\varepsilon_k) \ne 0$ uniformly distributed in the interval $[-\varepsilon,\varepsilon]$.
We observe numerically that
\begin{equation}
  \lim_{\varepsilon \to 0} \psi_N(x,0;\bs \lambda^\varepsilon) = \psi(x,0;\bs \lambda)
\end{equation}
---see Figure~\ref{fig:vto0}.
Note that this
limit is no longer valid for $t>0$ as solitons with ${\rm Re}(\lambda^\varepsilon_k)\neq 0$ scatter away. Using conservation in time of the Kruskal integrals, we obtain
\begin{align}
  A_j & \sim
  \lim_{\varepsilon \to 0} \left(\int_{-\infty}^{+\infty}
  p_j[\psi_N(x,0;\bs \lambda^\varepsilon)] \rmd x \right),\nonumber \\
      & = \lim_{\varepsilon \to 0} \left(\int_{-\infty}^{+\infty}
  p_j[\psi_N(x,t;\bs \lambda^\varepsilon)] \rmd x \right),
\end{align}
such that identities~\eqref{eq:p1}, \eqref{eq:p3} still hold for bound state SG.

We define now
\begin{equation}
  B_j = \int_{0}^{T} q_j[\psi(0,t)] \rmd t,
\end{equation}
where $T\gg 1$ such that
\begin{equation}
  \label{eq:aver2}
  \aver{q_j[\psi]} = \frac{B_j}{T} + O(T^{-1}).
\end{equation}
The average is evaluated at $x=0$ for
computational convenience only, and
the ergodicity property guarantees that this average
is uniform in space. We define the $N$-soliton solution $\psi_N(x,t;\bs \lambda)$ which approximates $\psi(x,t)$ in the temporal window $[0,T]$ at the position $x=0$ and cancels outside the window:
\begin{equation}
  \label{eq:window}
  \psi_N(0,t;\bs \lambda) \sim
  \begin{cases}
    \psi(0,t), & \text{if } 0  \le t \le T, \\
    0,         & \text{otherwise,}
  \end{cases}
\end{equation}
and similarly for the spatial derivatives $\partial_x \psi_N$.
By definition
of the flux $s(\lambda) f(\lambda)$, there are
$T s(\xi,\eta) f(\xi,\eta) |\rmd \lambda|$ right-propagating
solitons ($s(\lambda)>0$) with parameters
$\in [\lambda,\lambda+\rmd \lambda]$
that cross the position $x=0$ between the times $t=0$ and $t=T$; the corresponding number
of left-propagating solitons ($s(\lambda)<0$) is given by
$-T s(\xi,\eta) f(\xi,\eta) |\rmd \lambda|$. Thus the spectral parameters of the $N$-soliton solution are distributed with $T |s(\xi,\eta)| f(\xi,\eta)>0$.
Since $\psi_N(0,t;\bs \lambda)$ exponentially decays in the regions $t<0$ and $t>T$, we have
\begin{equation}
  \label{eq:Jn}
  B_j
  \sim \int_{-\infty}^{+\infty} q_j[\psi_N(0,t;\bs \lambda)] \rmd t.
\end{equation}
$\psi_N$ is only an approximation of the realization $\psi$ in $[0,T]$, especially at the boundaries $t=0$ and $t=T$ where $\psi_N(0,t;\bs \lambda)$ decays to $0$, and the correction to the integral~\eqref{eq:Jn} is $O(1)$.
The Kruskal integrals $\int_\mathbb{R} q_j[\psi_N(x,t;\bs \lambda)] \rmd t$
are now space-conserved quantities and
\begin{equation}
  B_j \sim \int_{-\infty}^{+\infty} q_j[\psi_N(x,t;\bs \lambda)] \rmd t,
\end{equation}
for any position $x$. In the limit $x \gg 1$, the
$N$-soliton solution can be approximated by a linear superposition of
distinct solitons, i.e. the solitons cross the position $x$ at very different time, and $B_j$ is given
by
\begin{equation}
  \label{eq:sumJ}
  B_j \sim \sum_{k=1}^N Q_j(\lambda_k),\quad
  Q_j(\lambda_k) =  \int
  q_j[\psis(x,t;\lambda_k)] \rmd t,
\end{equation}
where $\lambda_k$ are distributed by the spectral density $T |s(\lambda)|f(\lambda)$.
We have in particular:
\begin{equation}
  Q_2(\lambda_k) = -4|{\rm Im}(\lambda_k^2)|.
\end{equation}
The continuous limit of~\eqref{eq:sumJ} for a sufficiently large number
of solitons $N$ reads
\begin{equation}
  \label{eq:continuousJ}
  B_j \sim \int_{\Gamma^+} Q_j(\lambda) \, T |s(\lambda)| f(\lambda) \,
  |\rmd \lambda|.
\end{equation}
Substituting the integral~\eqref{eq:continuousJ} in the limit~\eqref{eq:aver2}, we obtain
\begin{align}
  \label{eq:q2}
   & \aver{q_2[\psi]}
  = 4 {\rm Im}\left(\int_{\Gamma^+} \lambda^2 s(\lambda) f(\lambda) |\rmd \lambda| \right),
\end{align}
where we assume the property ${\rm sgn}(s(\lambda)) = {\rm sgn}(-4\xi)$, with $-4\xi$ being the free soliton velocity, see \cite{congy_soliton_2021}.

\section{Partially coherent wave}

\subsection{Derivation of the DOS}

As indicated in the main text, the distribution of the discrete IST spectrum of a partially coherent wave for $x \in [0,L]$ is approximated by the Bohr-Sommerfeld density
\begin{align}
   & \varphi_L(\eta) = \int_I \frac{\eta}{\pi
  \sqrt{\rho_0(x) - \eta^2}} \,\rmd x, \nonumber                \\
   & \text{with: }I = \{ x\in [0,L] \;|\; \rho_0(x) > \eta^2\}.
\end{align}
The change of variable $\rho = \rho_0(x)$ yields
\begin{equation}
  \label{eq:PCWsum}
  \varphi_L(\eta) = \int_{\eta^2}^\infty \frac{\eta}{\pi
    \sqrt{\rho - \eta^2}} \sum_{i} |\rmd x_i|,
\end{equation}
where $x_i(\rho)$ is a solution of $\rho_0(x_i) = \rho$ and $|\rmd x_i| = |x_i'(\rho)| \rmd \rho$. Each interval of $\rho_0(x) \in [\rho,\rho+ \rmd \rho]$ makes the contribution $\eta/\pi \sqrt{\rho -\eta^2}$,  weighted by $\sum_i |\rmd x_i|$,  to the Riemann sum in ~\eqref{eq:PCWsum}.  Thus $\varphi_L(\eta)$ can be seen as a ``linear superposition'' of Weyl's distributions (Equation (10) in the main text). 

The total number of solitons ``contained'' in the portion $x \in [0,L]$, $L \gg \ell$, of  the partially coherent initial data is 
$N_{L}  \sim \int  \varphi_L(\eta) \rmd \eta$. The  
``global'' (i.e. defined for large spatial scales $\Delta x \sim O(L)$)
DOS corresponding to the partially coherent wave is then
given by
\begin{equation}
 f_0(\eta) \sim \frac{\varphi_L(\eta)}{L} = \int_{\eta^2}^\infty \frac{\eta}{\pi
    \sqrt{\rho - \eta^2}} \bigg( \frac{1}{L}\sum_{i} |\rmd x_i| \bigg)
\end{equation}
as $L \to \infty$.  The sum $\sum_{i} |\rmd x_i|/L$ corresponds to the probability $\PDF_0(\rho) \rmd \rho$ to have $\rho_0(x) \in [\rho,\rho+ \rmd \rho]$. 
Isospectrality of the fNLS evolution implies conservation of the global DOS so that after the generation of the homogeneous SG via the stochastization of soliton phases  at $t \to \infty$  the conventional, locally defined, DOS $\tilde f(\eta)=f_0(\eta)$ since for the homogeneous SG the local and the global DOS coincide. This yields formula (11) in the main text. 
We emphasize that the global DOS $f_0(\eta)$ (30) is defined for the partially coherent initial data, while the expression of the  kurtosis (Equation (7) in the main text)  is valid only for a homogeneous SG, i.e. in the long time regime.

\subsection{Relation between the PDF moments and the DOS moments}

The formula for the DOS of partially coherent wave (Equation (11) in the main text) yields a relation between the moments of the reduced DOS, $\overline{\eta^j} = \int_0^\infty \eta^j \tilde{f}(\eta) \rmd \eta$, and the moments of the PDF $\PDF_0(\rho)$ of the initial wave field. Consider $\overline{\eta^{2k+1}}$:
\begin{equation}
  \overline{\eta^{2k+1}} = \int_0^\infty \left(\int_{\eta^2}^\infty \frac{\eta^{2k+2}}{\pi \sqrt{\rho - \eta^2}} \PDF_0(\rho) \rmd \rho \right) \rmd \eta.
\end{equation}
The change of variable $\rho = \mu^2+\eta^2$ with $0<\mu<\infty$ yields
\begin{equation}
  \overline{\eta^{2k+1}} = \frac{1}{\pi}\int_0^\infty \int_0^\infty 2\eta^{2k+2} \PDF_0(\mu^2+\eta^2)\rmd \mu \rmd \eta .
\end{equation}
The further change to polar coordinates $\mu=\sqrt{\rho} \cos \theta$, $\eta=\sqrt{\rho} \sin \theta$ transform the integral
\begin{align}
  \overline{\eta^{2k+1}} & = \frac{1}{\pi} \bigg( \int_0^{\pi/2} \sin^{2k+2} \theta \rmd \theta \bigg) \bigg(\int_0^\infty \rho^{k+1} \PDF_0(\rho) \rmd \rho \bigg) ,\nonumber \\
  \label{eq:dos0}
                    & = \frac{(2k+2)!}{2\times 4^{k+1} [(k+1)!]^2} \aver{|\psi(x,0)|^{2k+2}}.
\end{align}
In particular we obtain $\aver{|\psi(x,0)|^2} = 4\overline{\eta}$ and $\aver{|\psi(x,0)|^4} = 16/3\overline{\eta^3}$.

\subsection{Alternative derivation for the kurtosis doubling}

The result  $\kappa_{\infty}=2 \kappa_0$  for partially coherent waves (Equation (12) of the main text) can be inferred directly from the definitions of the conserved quantities and fluxes (Equation (4) of the main text) and the basic assumption of the long-time resolution of a partially coherent wave via the bound state SG.

First we observe that the relation $\aver{q_2}=0$, valid for the bound state SG, implies  that for the long-time evolution of a partially coherent wave we have
\begin{equation}\label{app1}
  \aver{|\psi(x,t)|^4} =
  2\aver{|\psi_x(x,t)|^2}  \ \ \ \text{as} \ \ \ t \to \infty.
\end{equation}

Next, conservation of  $\aver{p_3(x,t)}$ in time implies that
\begin{align}
  &\aver{|\psi(x,t)|^4} -
  \aver{|\psi_x(x,t)|^2} \nonumber\\
  &\hspace*{1cm}= \aver{|\psi(x,0)|^4} -
  \aver{|\psi_x(x,0)|^2},
  \label{app2}
\end{align}
assuming ergodicity.

Since $|\psi_x(x,0)|^2 = \rho_0(x)([\rho_0'(x)/2\rho_0(x)]^2+u_0^2(x))$, we can assume that $ \aver{|\psi_x(x,0)|^2} \to 0$ in the semiclassical limit (cf. Equation (10) of the main text).
Thus, we obtain from~\eqref{app1},\eqref{app2} that $\aver{ |\psi(x,t \to \infty)|^4}   =   2 \aver{ |\psi(x,0)|^4 }$   and, using the conservation of
$\aver{|\psi|^2}$, we obtain
\begin{equation}\label{app3}
  \lim_{t\to \infty}\left(\frac{\aver{ |\psi(x,t)|^4}}{\aver{ |\psi(x,t)|^2}^2}\right)  = 2 \frac{\aver{ |\psi(x,0)|^4}}{\aver{ |\psi(x,0)|^2}^2}.  \end{equation}

\end{document}